# Advanced Electrical Conductors: An Overview and Prospects of Metal Nanocomposite and Nanocarbon Based Conductors

**Mehran Tehrani**

Walker Department of Mechanical Engineering, University of Texas at Austin, Austin, TX, United States

## Abstract

Advanced electrical conductors that outperform copper and aluminum can revolutionize our lives, enabling billions of dollars in energy savings and facilitating a transition to an electric mobility future. Nanocarbons (carbon nanotubes and graphene) present a unique opportunity for developing advanced conductors for electrical power, communications, electronics, and electric machines for the aerospace, marine, and automotive industries. This paper compiles the major research progress in the field of advanced nanocarbon-based and metal-nanocarbon conductors. It also elucidates the competitiveness and implications of advanced conductors with respect to conventional ones and describes their materials science, properties, and characterization. Finally, several areas of future research for advanced electrical conductors are proposed.

*Keywords: Carbon nanotube; graphene; copper; aluminum, conductor; electrical conductivity; covetic; ultra-conductive copper; advanced conductor*

Table of Contents

# 1   Introduction

Aluminum (Al), copper (Cu), gold (Au), and silver (Ag) are the most electrically conductive metals and therefore are used as the conductor material in different applications. Choosing the right conductor for an application requires cost and performance analyses, which in turn calls for a comprehensive knowledge of materials' electrical and thermal conductivities, mechanical performance, low and high temperature behavior, thermal expansion, corrosion resistance, etc. Cu and Al are the standard electrical conductors traditionally used for power generation, transmission, distribution, electrical equipment, electronics, and communications. This is due to their unrivaled combination of electrical conductivity, physical and mechanical properties, and cost. Silver outperforms both Cu and Al in terms of conductivity, however, its high cost and inferior mechanical performance has limited its applications. Gold is less conductive than Cu but is used in highly corrosive environments due to its exceptional corrosion resistance. In certain applications, conductivity may be compromised for strength, corrosion resistance, low thermal expansion, etc. Over the last two decades, the need for power dense and high efficiency electrical systems has resulted in the emergence of advanced conductors with unique sets of properties.

Room temperature advanced conductors can be categorized into nanocarbon-based conductors and metal-carbon conductors, as shown in **Figure 1**. Allotropes of carbon such as graphite, graphene (Gr), and carbon nanotubes (CNT) constitute the basis for both nanocarbon-based and metal-nanocarbon nanocomposite conductors. Nanocarbon-based conductors are comprised mostly of CNT or graphene with small quantities of other elements and compounds for improving their conductivity. Metal-carbon conductors are metals that contain a small loading of CNT, graphene, or in rare cases other carbon allotropes. These conductors can potentially offer advantageous properties over pure metal conductors, such as a lighter weight, a higher current carrying capacity, or survivability in mechanically or chemically harsh environments.[1]

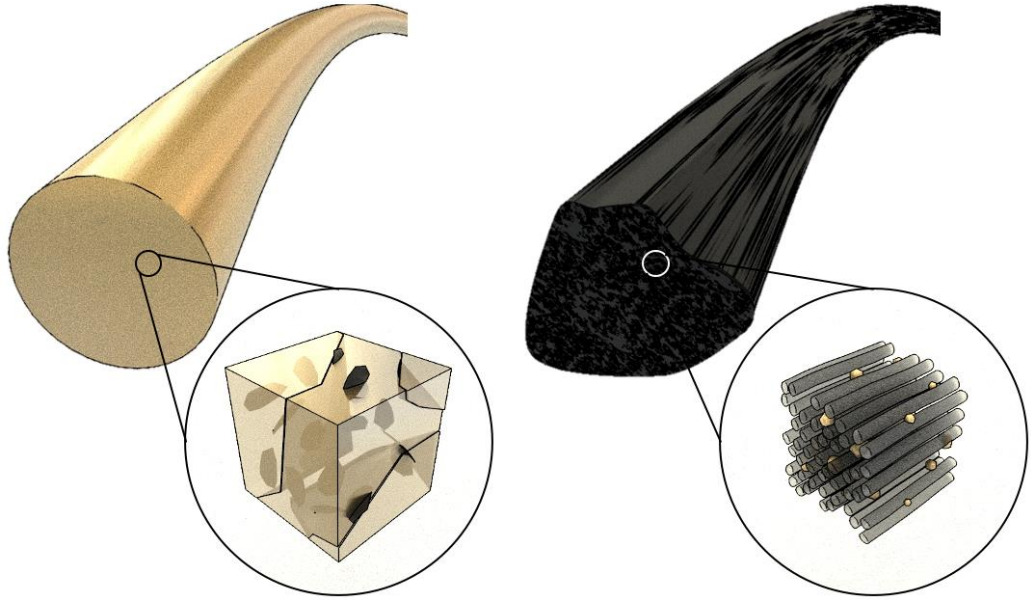

*Figure 1. Schemtics of advanced electrical conductors: metal-nanocarbon (left figure: metal matrix enhanced with nanocarbons) and nanocarbon-based (right figure: nanocarbon matrix enhanced with dopants such as metal nanoparticles, metal-halides, acids, etc.).*

Electrical conductors are required for many areas of technology that affect our lives, from sensor wires to motor windings and power transmission lines, and from electromagnetic interference (EMI) meshes to high voltage bus bars. Over the last two decades, there has been a plethora of publications and patents around advanced electrical conductors. This paper compiles the major findings related to these materials and places them in the perspective of industrial relevance; i.e., whether the measured properties, usually laboratory scale, are reliable and can be achieved at the industrial scale, and if these new conductors can be manufactured in a scalable and economical approach. Additionally, the paper scrutinizes the competitiveness and implications of advanced conductors with respect to Cu and Al and describes the correlations between the microstructure and several properties of interest in advanced conductors.

The price of advanced conductors is currently higher than conventional metal conductors, therefore, their performance should be considerably better to justify their use. Even with this expensive price tag advance conductors have found a niche in the aerospace and defense applications. It should be noted that it almost took a century and several technological breakthroughs to bring copper conductors to fruition.[2] Given the market need for superior conductors, it is reasonable to presume that it is only a matter of time until innovations in materials science and manufacturing reduce the price and enhance the performance of advanced electrical conductors, making them ubiquitous by replacing Cu and Al.

## 2 Electrical Conductors

Before delving further, the baseline properties of standard conductors (Cu and Al) should be highlighted (see **Table 1**). Copper wires (cold drawn) have a yield strength of ~300 MPa and a density of 8.9 g/cm$^3$. The performance of conductors is usually normalized with respect to the International Annealed Copper Standard (IACS), which has an electrical conductivity of 58.1 MS/m at 20°C. For example, the conductivity of aluminum alloy 1350, which is used widely in electrical applications, is ~62% IACS and the conductivity of most commercial copper wires is 101% IACS. Aluminum conductors have a yield strength of ~110 MPa and density of 2.8 g/cm$^3$. Aluminum has a higher coefficient of thermal expansion (CTE) compared with copper (23 vs.17 /μ/K), creeps at connections where temperature may locally raise, and is prone to galvanic corrosion in humid environments. Aluminum is in general vulnerable to oxidation and corrosion at connections. The oxidation/corrosion results in high electrical resistances leading to local hot spots, melting the insulation or fixture and causing fires in some cases. Copper offers better bending characteristics for installations and handling. Copper is, therefore, more accepted and often used in weight critical applications; despite its 50% lower conductivity per weight compared with aluminum. Finally, the ampacity (current carrying capacity) of copper is higher than an aluminum wire of the same diameter, making it the material of choice in generators, transformers, and electric motors where a compact design is usually required; ampacity is the maximum current a conductor can carry without surpassing its temperature rating.

### 2.1 Carbonaceous conductors

As shown in **Figure 2**, there are different electrically conductive allotropes of carbon, namely "carbon nanotube", "graphene", and "graphite". Graphene is a single layer SP$^2$ bonded hexagonal network of carbon atoms and graphite is a stack of several graphene layers. Graphene, however, has unique electronic properties that are distinctly different than those of bulk graphite. These unique characteristics are also preserved when only a few layers of graphene are stacked.

Once the number of stacked layers passes 4-5, electronic properties of the multi-layered graphene approach bulk graphite and the term "graphene" should no longer be used. In the literature, however, graphite nanoplatelets comprising tens of graphene layers may sometimes be referred to as "graphene". Highly functionalized graphene (graphene oxide, GO, or reduced graphene oxide, rGO) with physical and mechanical properties far from those of pristine graphene are unwittingly referred to as graphene in the literature. The term "carbon nanotube" refers to variations of tubular nanocarbons with distinct qualities. The major categories are single-walled (SWCNT), double-walled (DWCNT), and multi-walled (MWCNT) CNTs.[3] A nanotube can be imagined as a seamless tube made from a rolled graphene, as shown in **Figure 2**. CNTs display a metallic or semiconducting behavior depending on this rolling angle. CNT or graphene used to make conductive wires and electrodes are usually not the ideal defect and impurity-free structures shown in **Figure 2** but rather contain impurities and structural defects. For CNTs, the number of walls significantly impacts their electrical conductivity and quality, with their single and double walled varieties exhibiting the highest quality and conductivity.[4] Unfortunately, the price of nanocarbons increases exponentially with their purity and quality.

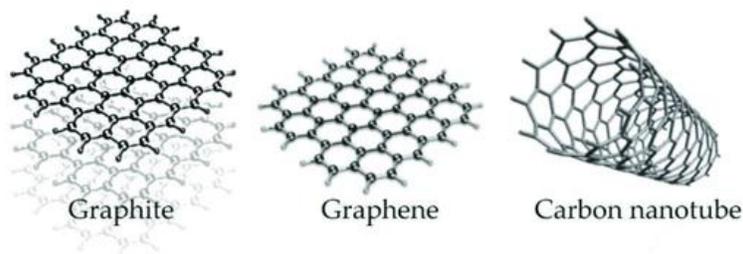

*Figure 2. Electrically conductive allotropes of carbon.[5]*

Since the first observation of a SWCNT in 1993 and discovery of graphene in 2004, nanocarbons have attracted great interest. An individual nanocarbon (e.g., carbon nanotube or graphene) exhibits a unique and extraordinary set of mechanical and physical properties as summarized in **Table 1**.[6] Numerous applications have been envisioned and explored for these nanomaterials, both theoretically and experimentally. While individual nanotubes and monolayer graphene exhibit outstanding properties, performance of their micro- and macro-scale assemblies depart greatly from the ones at the individual level due to particle-particle interactions, defects, and impurities. Another reason for such discrepancy between the macro- and nano-scale properties may be that properties of individual CNT or graphene are usually measured for a free-standing sample in vacuum, which is different than the realistic case of a macro-scale sample examined in air where each nanocarbon is interacting with several other particles. An ideally packed structure of aligned finite length CNTs would be somewhat similar to a nanocrystalline metal with grains that are microns long and nanometers in diameter, scattering charge carriers at its numerous grain boundaries; the mean free path of electrons in an isolated nanocarbon is orders of magnitude larger than that of bulk conductive metals.[7] Boundary scattering can be theoretically resolved if ultra-long (or continuous) and defect-free nanocarbons are electronically separated in the transverse direction and aligned in the wire direction.[8] Achieving such structure has proven to be extremely challenging.

There has been tremendous progress toward producing CNT and graphene fibers (aligned assemblies) in order to translate individual nanocarbon properties to the macro scale.[1b, 9] The highest reported electrical conductivities for pure CNT and graphene fibers are two orders of magnitude lower than copper. Dopants (strong acids, metal halides, halogens, etc.) are therefore incorporated in nanocarbon fibers to enhance their charge carrier density and reduce the nanocarbon-nanocarbon junction resistances.[10] The current record conductivity for doped carbon nanotube fibers is 8.5MS/m (~15%IACS)[4] and 1.0MS/m for a graphene fiber[11]; potassium doped graphene fibers (unstable in air) have demonstrated a superb electrical conductivity of 22.4MS/m (~38%IACS).[12]

While nanocarbon fibers exhibit electrical conductivities that are lower than copper's, the low density of these fibers renders a higher specific conductivity for the doped CNT and graphene fibers than copper wires.[4, 12]; specific conductivity is conductivity divided by density, i.e., conductivity normalized by weight. Stability and toxicity of some of the most effective dopants, in addition to their high price, are concerning for their potential applications. **Table 1** tabulates the highest reported properties for bulk carbonaceous fibers and conventional metals.

*Table 1. The best reported properties (ambient condition) for conventional and advanced electrical conductors. Data in this table is provided for comparison purposes only. IACS is conductivity over that of international annealed copper standard.*

|  | IACS% | Strength (GPa) | Density (g/cm$^3$) |
|---|---|---|---|
| Individual CNT or Graphene | 33-172 | 20-100 | 1.4 |
| Doped CNT fiber[4] | 15 | 3 | 1.5 |
| Doped graphene fiber[11] | 2* | 2 |  |
| Doped carbon fiber[13] | 24 | 1 | 2.5 |
| Cu-CNT nanocomposite[14] | 47** | - | 5.2 |
| Ultra-conductive copper[15] | 117 | - | 8.9 |
| Copper | 101 | 0.3 | 8.9 |
| Aluminum | 62 | 0.1 | 2.7 |
| Silver | 105 | 0.1 | 10.5 |

*an electrical conductivity of 38% IACS has been reported for a potassium-doped (measured in inert atmospheres) graphene fiber.[12]

**exhibits a better electrical conductivity than copper above 80 °C.[14]

Carbon fibers have been around for over 50 years and exhibit mechanical and physical properties that are on par with nanocarbon-based fibers. As shown in **Figure 3**, carbon fibers are made from polymer precursors such as cellulose, polyacrylonitrile (PAN), and pitch. These polymers are spun into a fiber form and subsequently oxidized, pyrolyzed, and graphitized, respectively, yielding graphitic nanodomains and turbostratic carbon that are covalently bonded. CNT and graphene fibers, on the other hand, are made from CNTs and graphene particles, respectively, that are assembled into a macroscale form. Properties of all carbonaceous fibers improve with the order and crystallinity of their constituents. For example, large and high-quality graphene particles that are tightly packed and de-wrinkled result in enhanced electrical conductivity in graphene fibers.[16] Similarly, high quality CNTs that are packed and aligned in the fiber direction achieve better electrical conductivities.[8]

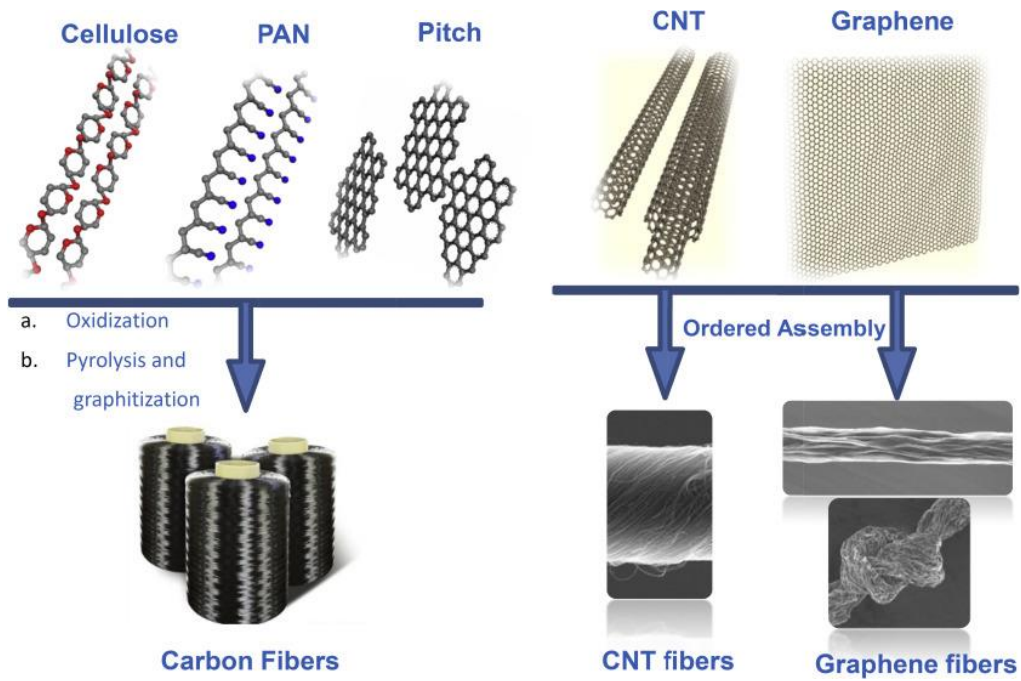

*Figure 3. Carbonacous fibers made form different precursors.[17] Polymer fibers undergo several post-processing steps to achieve carbon fibers while CNT and graphene fibers are made from CNT and graphene particles, respectively. Reproduced with permission from Elsevier.*

### 2.1.1 Carbon Nanotube Fibers

CNTs are usually inhomogeneous mixtures of tubes with various electronic properties and diameters. The ratio is usually one-third metallic and two-third semiconducting nanotubes; multi-walled CNTs (MWCNT) are comprised of nanotubes with different electronic properties. An ideal advanced conductor would consist of infinitely long metallic CNTs that run parallel to each other.[8] Although metallic CNTs are desired for their high electrical conductivity, scalable and cost-efficient methods to produce them do not currently exist. In reality, high-quality CNTs can be only synthesized to a finite length (usually tens of microns) and are mixtures of both semiconducting and metallic CNTs. A detailed discussion on the synthesis and processing of different CNT types can be found elsewhere.[3]

As summarized in **Table 2** and shown in **Figure 4**, three major routes for fabricating CNT fibers have been developed:[1a]

- The first is a dry processing approach to spin fibers directly from a pre-grown vertically aligned MWCNT array;[18] **Figure 4a**. These nanotubes are of a relatively low quality (based on their Raman $I_G/I_D$). These CNT fibers are >99% pure and achieve conductivities below 0.1MS/m.[19] These fibers are commercially available from Lintec Nano Science & Technology Center.

- The second approach is dry spinning of a CNT aerogel that forms downstream of a floating catalyst CVD (FC-CVD) reactor.; **Figure 4b**. CNT quality and number of walls can be optimized in FC-CVD.[20] Millimeters long single and few-walled CNTs (relatively medium to high quality) are directly spun into a fiber form. These fibers are usually subjected to post-process acid doping and densification, resulting in relatively high conductivities, 0.1-0.5MS/m and up to ~2MS/m when doped.[21] Catalysts and promoters used in the synthesis remain in the CNT yarns (a few% weight) as impurities. Amorphous carbon also forms between the CNTs and their bundles, bridging them and improving both their mechanical and electrical properties.[21b, 22] These fibers are commercially available from Huntsman (formerly Nanocomp Technologies Inc.).

- The third CNT fiber synthesis route is a wet processing approach that involves extruding a premade CNT-acid solution through a spinneret into a coagulation bath, resulting in dense CNT fibers;[23] **Figure 4c**. CNTs used in the wet processing route are of the highest quality and purity, and are up to twenty microns long.[4, 24] The strong acids used in this process decrease CNT-CNT junction resistance and dope the individual nanotubes, improving the conductivity of the CNT fiber by almost an order of magnitude to 8.5 MS/m (maximum of 15% IACS) for only a fraction of copper density.[4] These fibers are commercially available from DexMat.

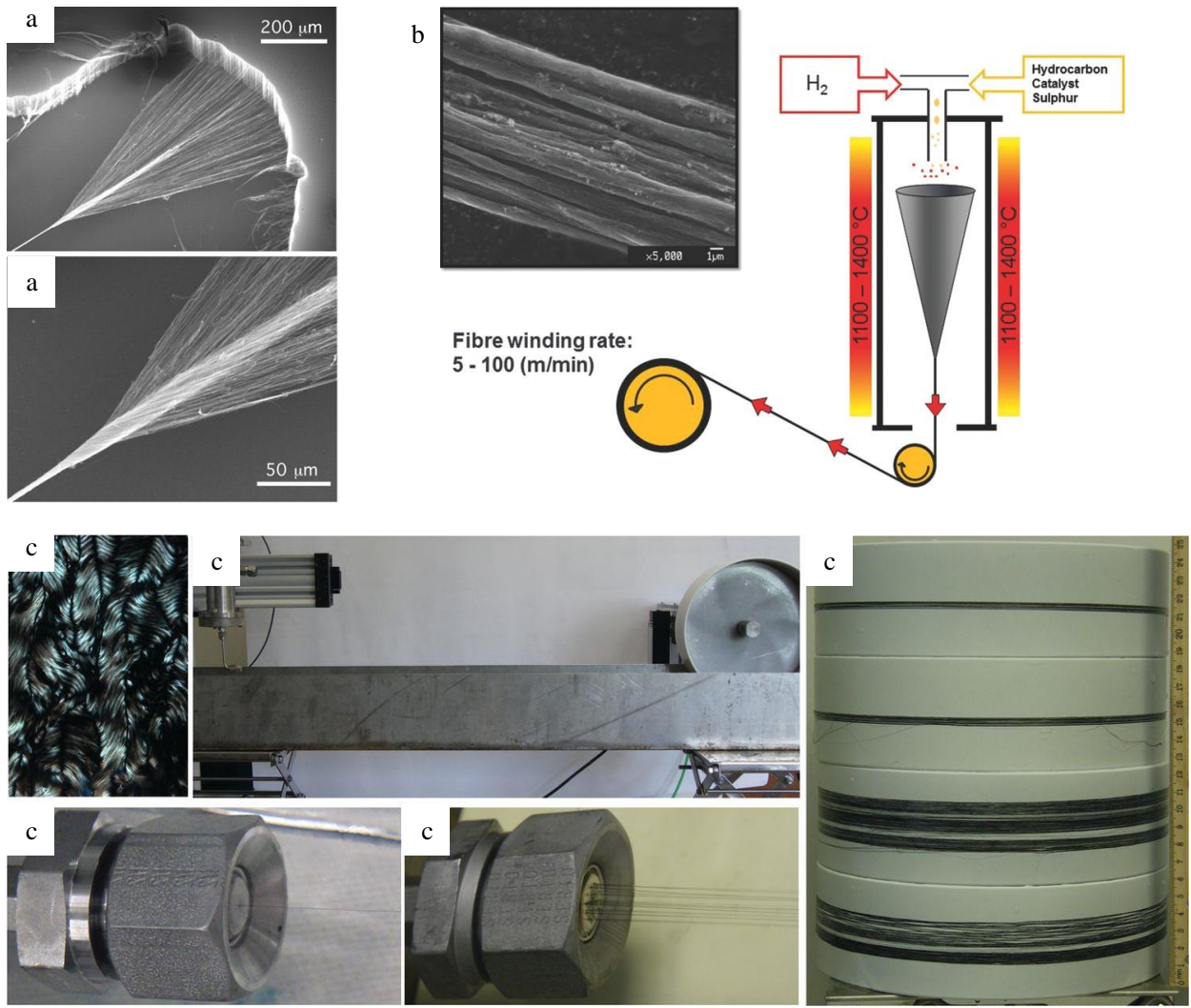

*Figure 4. Major processes for making CNT fibers: a) micrograph of spinning from a CNT forest,[25] b) schematic of spinning from a floating catalyst-CVD reactor and micrograph of a CNT yarn surface,[1b] and c) wet spinning from a CNT solution through single and multiple spinnerets.[26] Reproduced with permission from the American Association for the Advancement of Science and Wiley.*

Novel approaches for fabricating CNT conductor wires have been also investigated.[27] In general, electrical conductivity of CNT fibers can be improved by up to 10 times via acid,[28] iodine,[29] or metal nanoparticles[21a] doping. Doped nanocarbon fibers contain only a few weight percentages of a metal or metal halide,[12] in contrast to metal-carbon nanocomposites that are largely metal-based.

*Table 2. Commercially available CNT yarns and their properties. FW-CNT: few-walled CNT.*

|  | Dominant CNT Type | CNT length (μm) | Raman $I_G/I_D$ | Tensile Strength (GPa) | Electrical Conductivity (MS/m) | Density (g/cm$^3$) |
|---|---|---|---|---|---|---|
| Lintec: Dry spun from a CNT forest[1d] | MW-CNT | <500 | 1 | <1 | 0.1 | 0.5 |
| Huntsman: Dry spun from FC-CVD[30] | FW-CNT | <2000 | 2-5 | 1 | 0.3-2 | 0.5-0.9 |
| DexMat: Wet spun from a CNT-acid solution[31] | DW-CNT | <20 | >50 | 0.4-2.8 | 3-10 | 0.8-1.6 |

The dominant route toward CNT production is chemical vapor deposition (CVD), offering simplicity, low cost, and high yields[32]. Currently almost all large-scale CNT suppliers use CVD.[33] Ultra-high quality ($I_G/I_D$ ratio >50) but low-purity CVD grown carbon nanotubes are currently sold for ~$1000/lb[34]. There are of course additional post-processing steps to remove undesired impurities, dope CNTs, and form conductor wires. These steps can increase the final price of a CNT wire, however, their price can significantly decrease if demand is increased and synthesis reactors are scaled up. Lifetime electricity savings may also justify higher costs. Finally, the price of advanced conductors should be compared with that of Cu wires, where although the cost of scrap copper is relatively low, price of copper wires increases drastically with decreasing the wire diameter; it is somewhat difficult to compare the price of different conductors since comprehensive data on the performance of advancec conductors at various current/voltage/temperature ratings are not available.

### 2.1.2 Graphene Fibers and Films

Mono-layer, defect-free graphene has the highest electrical, thermal, and mechanical properties ever reported.[35] Such high-quality graphene has been reported to surpass copper's electrical conductivity but can be only grown via CVD. CVD synthesis of graphene is suitable for making transparent and ultra-conductive surface coatings but maybe not yet economical for manufacturing macroscale conductors. Graphene films and fibers are instead made from relatively low-quality graphene in the form of few-layered graphene oxide. In contrast to CNTs, graphene oxide (GO) can be processed in water and its low cost makes GO structures highly attractive for several applications. For electrical conductors, intensive post-processing of GO in reducing environments and extreme temperatures (up to 3000 ºC[9b]) is required to achieve reduced graphene oxide (rGO) with an enhanced quality, close to that of pristine graphene. As shown in **Figure 5a**, graphene oxide is made by oxidizing and exfoliating graphite.

Mechanical and electrical properties of rGO fibers[9b] and films[36], shown in **Figure 5b** and **c**, are usually lower than the best values reported for CNT fibers. For example, graphene conductors (films and fibers) exhibit electrical conductivities and tensile strengths up to 0.1 MS/m and 1GPa, respectively. In contrast, thermal conductivity of graphene films and fibers are comparable to or better than that of CNT fibers, achieving values as high as 2000 W/m/K[37]; individual CNT or Gr possess thermal conductivities of up to 5000 W/m/K. Larger, better-quality (lower defect), monolayer, and un-wrinkled graphene particles give rise to better properties in their macro structures.[37] A great deal of research has, therefore, been focused on the synthesis of large-sized graphene particles and their assembly into fiber and film forms.[9b] As mentioned earlier, these particles are usually oxidized during their synthesis and are subsequently reduced (their oxygen is removed) to recover some of their properties.

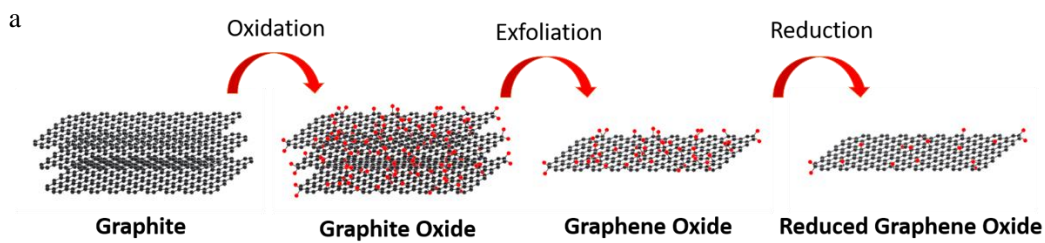

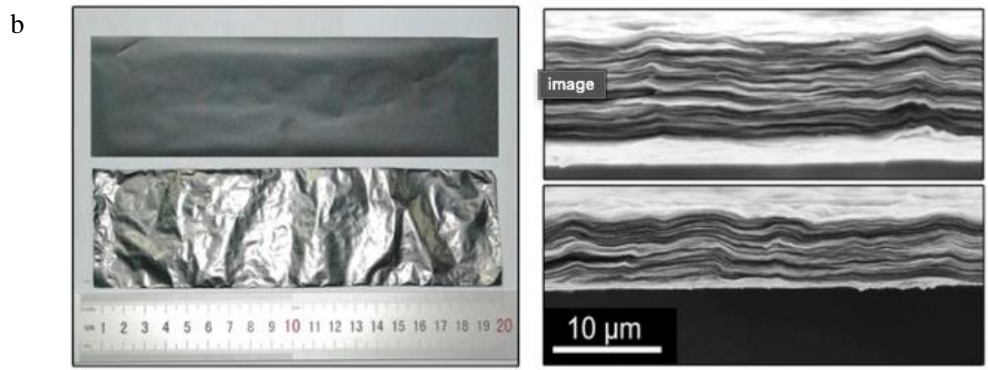

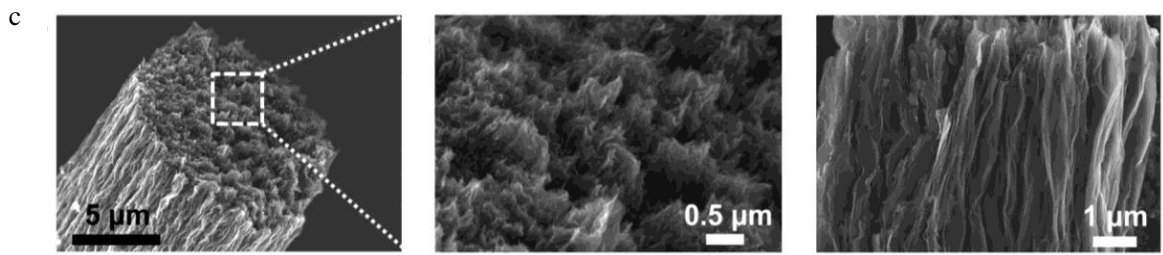

*Figure 5. a) Exfoliation of graphite to GO and reduction of GO to rGO; red dots represent oxygen-containing functional groups. b) Top (left) and side (right) view of a rGO film.[38] c) Structure of a rGO fiber at different length scales.[39] Reproduced with permission from the American Association for the Advancement of Science and Wiley.*

The challenge in this research field remains to be the discovery of a scalable process that retains properties of graphene across the scales to macroscale wires. CNT and graphene fibers need to be heavily doped with acids, metal, or metal halides to compete with Cu and Al. In consequence, over the last decade research in this field has partially diverted to understanding the processing and materials science of metal-nanocarbon hybrids.

## 2.2 Metal-Carbon Nanocomposites

It is important to distinguish between traditional composite, nanocomposite, and electronically-hybridized nanocomposite conductors. I) Composite: A composite conductor is made by combining two materials to form a new one that benefits from the properties of its individual constituents. The discontinuous phase is usually referred to as the 'additive' and the continuous one as the 'matrix'. As a 'hypothetical' example, copper (matrix: conductor phase) and carbon fibers (additive: structural phase) can be combined to achieve both high mechanical and electrical properties. The conductivity of the composite will be lower than that of the pure copper (but higher than carbon fiber's) and its mechanical properties lesser than those of carbon fibers (but larger than copper's). The ratio of fibers to copper can be modified to achieve a set of desired properties; a stronger composite has to have a higher carbon fiber loading and a more conductive one a lower carbon fiber loading. Upper and lower limits exist for composite properties, as demonstrated in **Figure 6**. II) Nanocomposite: A nanocomposite is similar to a composite with the distinction of a nanoscale additive phase; a nanomaterial is smaller than 100nm in at least one of its dimensions. The nano-scale additives and their high surface area may contribute to additional property improvement over traditional composites, specifically if their interfaces with the metal matrix are properly engineered. Nanomaterials such as CNT or graphene offer exceptional properties that can be exploited in a nanocomposite, however, their high surface area makes their handling and processing much more difficult than micro-scale additives used in composites. III) Electronically Hybridized Nanocomposite: If the nanocarbon and metal interact in a manner where their electronic band structures are modified in the presence of each other, an electronically hybridized nanocomposite is achieved; This has been the main motivation behind a metal-carbon conductor, combining the high electron mobility of nanocarbons with the high charge carrier density of metals to achieve unprecedented electrical conductivities; this can occur via chemisorption or physisorption between the nanocarbon and metal.[40] Improving the room-temperature electrical conductivity of Cu and Al via the integration of nanocarbons has proven difficult.

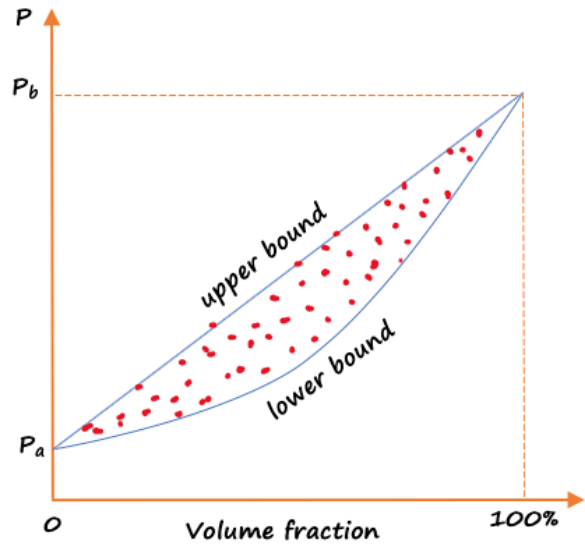

*Figure 6. Upper and lower bounds predicted by the rule of mixture for a mixture of "a" and "b" materials as a function of their volume fraction ($V_f$). Red dots designate expereimental values. P is the property of interest, and $P_a$ and $P_b$ represent the inherent property of materials a and b, respectively. Upper bound $P = P_a \times V_{fa} + P_b \times V_{fb}$ and lower bound $P = 1/( V_{fa}/P_a + V_{fb}/P_b)$.*

One of advantage of nanocarbons is their high aspect ratio and accessible surface area. While a benefit, a high surface area means a high surface energy that should be overcome for dispersing nanocarbons. For example, pristine CNTs are always in the form of agglomerates and require kilojoules of energy to de-agglomerate their milligram quantities and strategies to keep them separated. In general, even if nanocarbon interactions with the host matrix are weak, the combined effect of these individual interactions can potentially be enormous due to the nanocarbon's extremely large surface area. If these interactions are strong, from either a mechanical/thermal or electrical standpoint, superior properties may arise. The level of property improvement in metal-carbon hybrids depends on several factors including noncarbon dispersion quality, type of metal, quality of nanocarbon, and interfacial phenomenon at the nanocarbon-metal interface. The following cases for mechanical, electrical, and thermal properties of metal-carbon nanocomposites are provided to highlight the potential benefits of nanocarbon addition to metals:

- *Mechanical:* Metals can be strengthened via several mechanisms including "precipitation strengthening", "grain size reduction", and "alloying". The addition of small amounts of CNT or graphene (it can be in the form of GO or rGO) to a metal can improve its yield and ultimate strengths via the similar strengthening mechanisms known for conventional metals.[41] In particular, nanocarbons can act as barriers to dislocation movement in metal-carbon nanocomposites (similar to precipitate strengthened metals) and inhibit grain growth (similar to strengthening via grain size reduction). Strengthening through the alloying mechanism requires the dissolution of carbon in metals, which is not significant given the extremely low solubility of carbon in copper and aluminum. In addition to the aforementioned mechanisms, if the nanocarbon loading is more than only a few percentages, their strength and stiffness can directly contribute to the mechanical properties of the nanocomposite; e.g., a porous CNT fiber electroplated with copper has been demonstrated to achieve strengths in excess of 600 MPa, which is almost double that of pure copper.[42]
- *Electrical:* It is well-known that impurities in a metal scatter electron, reducing its electron mean free paths and degrading electrical conductivity. Nanocarbons in a copper or aluminum matrix act as impurities and usually degrade the host metal's conductivity; a few exceptions exist.[43] Similarly, while the mean free path of charge carriers in a freestanding CNT or graphene particle is very large, embedding that same nanoparticle in a metal may reduce this mean free path, reducing the mobility of electrons in the noncarbon phase. Other interesting phenomena may happen at the interface of the nanocarbon and metal. It is not yet completely understood how adding nanocarbons to a metal can improve the overall conductivity of the metal, however, it has been shown that electrons can easily transfer from copper to graphene. If transferred, electrons may travel at a faster velocity in the embedded graphene or Cu-graphene interfaces than they would in pure copper, improving the electrical conductivity.[15]

- *Thermal:* The thermal conductivity of a nanocarbon-metal nanocomposite can be calculated as a weighted sum of the nanocarbon conductivity (considering its orientation), metal conductivity, and the thermal resistance between the nanocarbon particles and the metal matrix, i.e., the Kapitza resistance. It turns out that it is difficult to disperse and align nanocarbons (recall the properties of nanotubes are the highest about their major axis and graphene about its 2D plane) in metal matrices and reducing Kapitza resistance is non-trivial. While the theoretical promise for improving properties is high, the reality is somewhat different. Thermal conductivity of metal-carbon nanocomposites is, therefore, limited to a maximum of 2000 W/mK.

In addition to these three major properties, the incorporation of nanocarbons in a metal matrix can improve several properties that are critical for a range of applications. For example, nanocarbon-copper composites can potentially outperform copper in terms of maximum current carrying capacity,[44] lower temperature coefficient of resistance,[14] and lower coefficient of thermal expansion.[45] The main challenge facing this field is to scale up and optimize the current fabrication technologies that are mostly developed on a lab scale.

Metal-carbon nanocomposites can be categorized based on their synthesis route into melt-processed and solid-state processed. Example of the former is "covetic" processing where a carbon precursor is added to a metal melt and subjected to high currents. Solid-state processes for making metal-carbon conductors, such as electrodeposition of copper on a CNT fiber,[46] are more widely used as they require simpler setups and lower input energy.

### 2.2.1 Melt-Processed Metal-Carbon Nanocomposites

Covetics are an interesting category of metal-carbon nanocomposites that have been around for more than a decade now. Despite superb electrical conductivities reported for nanoscale covetic films,[47] researchers have failed to produce bulk covetic samples that outperform copper's electrical conductivity. The microstructure of a covetic metal can be different than conventionally processed ones. The exotic microstructure of covetics has also not yet translated to improved electrical, thermal, or mechanical improvements in bulk covetic samples. Research on covetic materials has been limited to several groups with the knowledge of making these materials. The processing of covetics is largely unknown and their microstructure and materials science is not well underestood.[48]

To make covetics, a carbon source is added and vigorously mixed with a molten metal (usually copper, aluminum, or silver) while a high electrical current is applied to the mixture, partially converting the precursor carbon into nanoscale carbon forms such as graphene nanoribons.[49] The carbon source is usually added in stages while the melt is stirred under high electrical currents for minutes.[50] Covetics exhibit a microstructure containing both micro-scale non-converted carbon and nanocarbon layers in between the metal atoms. It is often stated that covetics contain relatively large amounts of carbon (>8 wt.%) in Al, Cu, and Ag.[51] It is, however, not clear how much of the precursor carbon has been converted during the covetic processing and how much is remained unchanged. Unfortunately, techniques that measure the amount of converted carbon can only examine extremely small regions of the samples. Regardless of what the exact fraction of converted carbon is, the resulting material does not exhibit an enhancement in properties of interest, i.e., electrical and thermal conductivities. Another major hurdle with covetics processing is achieving uniform carbon distribution in the host metal. Mixing technique, therefore, has a significant effect on the structure and properties of covetics.

The existence of carbon in the graphene form is confirmed in covetics.[52] Graphene can be grown on certain metals (and molten metals[53]) including nickel and copper. To this end, chemical vapor deposition (CVD) is used. CVD is a well-established technique for graphene growth where a hydrocarbon, the carbon source, is first decomposed to carbon at high temperatures on the metal surface and subsequently dissolves in it. A supersaturated metal-carbon solution forms and carbon is precipitated in the form of graphene on the surface. Similarly, graphite powder or chunks can be used as the carbon source and graphene can be grown in a metal melt under high electrical currents.[53] This is possibly one scenario under which carbon nanostructures form inside covetics, i.e., form on the surface and are subsequently incorporated into the metal under mechanical mixing. It is, however, unclear how epitaxial layered structures of metal-carbon are formed in covetics.

Covetic properties are highly dependent on the carbon precursor material and processing conditions, where some of the carbon particles in a covetic metal may act as a defect degrading the properties and others may improve certain properties.[50] As mentioned earlier, there is only a little that we know about covetics. Local hot spots, reaching thousands of degrees of Kelvin, might form within the melt due to electrical arcing and carbon ionization. This might be the unique aspect of covetic processing,[52] allowing for conversion of carbon and its incorporation into the metal lattice in 'far-from-equilibrium' conditions.

The addition of covetic carbon changes the micro-structure of the metal, and therefore, alters its mechanical performance. The resulting material demonstrates a higher yield and ultimate strength.[54] In the bulk form, electrical conductivity of covetics is similar to the pure metal. Such conductivity is considered high given the high level of impurities (Fe, Ni, and S) that exist in covetic metals.[50] These impurities degrade the conductivity of copper under normal circumstances; the implication here is that the covetics fabrication process should be greatly controlled and use highly pure precursors to prevent contamination. Thin covetic films (~30 nm) fabricated using physical vapor deposition (PVD) have exhibited up to 130% IACS. While this result is highly intriguing, it is not clear whether the PVD is responsible for forming the high conductivity film or if such conductivities exist within the covetic metal but only in certain regions. There are few reports, from 2012, on superior bulk electrical and thermal properties of covetics. However, the reported superior properties of the best samples have never been reproduced. The reader should, therefore, be cautious about the promise of covetics.

### 2.2.2 Solid-State Processed Metal-Carbon Nanocomposites

Most published studies on metal-carbon conductors report degradation or no change in electrical conductivity with the addition of nanocarbons to metals. Some studies exhibit improvements over a reference sample, but their measured conductivities are still inferior to commercial materials such as the IACS. There aren't, therefore, many investigations that have published an enhanced electrical conductivity in metal nanocarbon nanocomposites with reference to the commercial copper or aluminum wires. These success stories are summarized here, highlighting their differentiating factor and implications for real-world applications. The common manufacturing techniques for metal-carbon nanocomposites include powder metallurgy, melting/solidification, spark plasma sintering (SPS), and electrodeposition. An example of a hybrid copper-graphene composite wire made by electrodepositing copper on a CVD graphene fiber is shown in **Figure 7**.

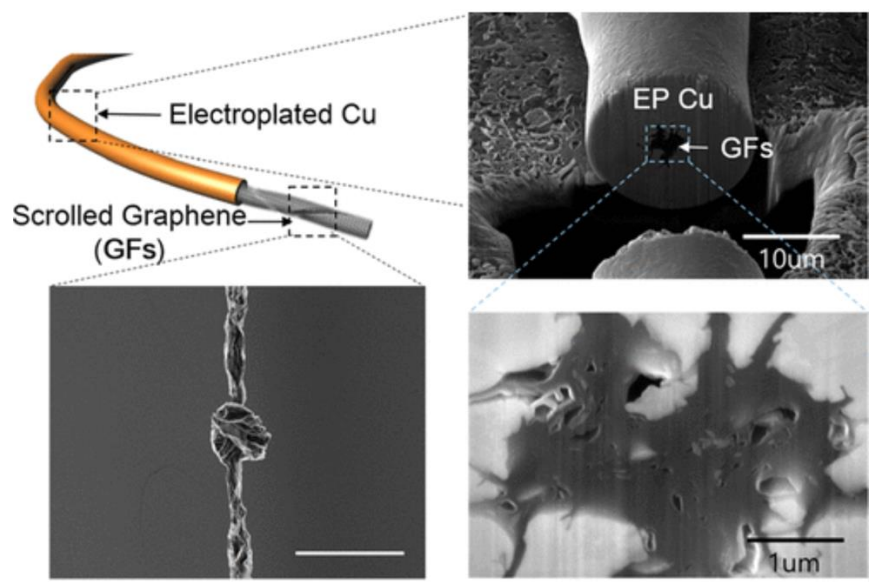

*Figure 7. A copper-shell graphene-core fiber. Reproduced with permission from the American Chemical Society.* [55]

The National Institute of Advanced Industrial Science and Technology (AIST), Japan, has published a series of studies that have had a significant effect on the R&D of advanced conductors.[56] In one of their first studies in 2013, AIST researchers showed interesting properties for a CNT-Cu nanocomposite. To fabricate their samples, they used a rolled-down vertically grown SWCNT and a multi-step process to ensure thorough impregnation of the porous CNT network with copper. To this end, a two-step copper electrodeposition using organic and inorganic solutions is used; with several reducing/annealing steps in between to remove any oxides formed. Their centimeter-long Cu-CNT sample achieves a 42% lower density and a better electrical conductivity than IACS above ~80°C, as shown in **Figure 8**. Their nanocomposite sample also exhibited a 100x improvement in maximum current carrying capacity in high vacuum (1e-6 torr) over copper. This improvement was attributed to the increase in copper's diffusion activation energy due to CNT presence. Specifically, copper's low activation energy at surface and grain boundaries was much improved due to CNT presence by suppressing the Cu diffusion pathways. This study suggests that CNTs (or potentially any nanocarbon) can be incorporated at grain boundaries and outer surfaces of a conductor to suppress electromigration and lead to an improved maximum current rating.

In a follow-up study in 2016, AIST exhibited the scalability/feasibility of this approach for microscale electronics and inverters by patterning SWCNTs on a silicon wafer and fabricating Cu-CNT micro-patterns.[44] The performance of the micro-traces was close to the macroscale films. In addition to the improved high current carrying capacity in vacuum, the Cu-CNT conductive traces showed a relatively low CTE, matching that of the silicon substrates. The Cu-CNT, therefore, offers a solution to micro-electronics failure due to thermal cycling.

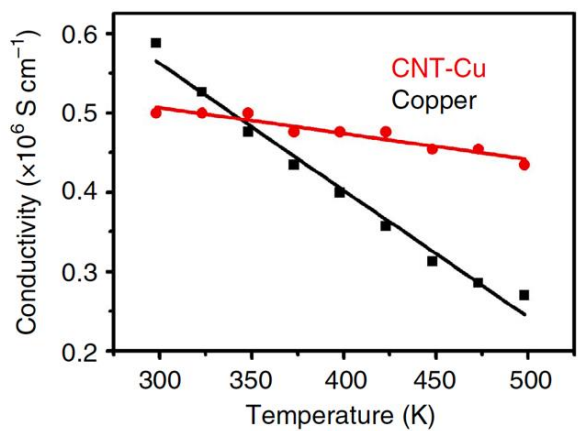

*Figure 8. Centimeter scale CNT-Cu film exhibiting a better electrical conductivity than copper above 80ºC. Reproduced with permission from Springer.[14]*

Given the success of small-scale Cu-CNT films, several groups in Europe, Asia, and the U.S. investigated the electrodeposition of copper on CNT and graphene yarns to achieve an elevated temperature ultra-conducting Cu-CNT wire.[14, 46, 57] Multi-walled CNT yarns have been usually used in these studies due to the difficulty of making macroscale and porous SWCNT fibers. To improve the wetting of hydrophobic CNTs with electrolyte, two approaches have been mainly reported: i) CNT functionalization with oxygen-containing groups and ii) using an organic solvent, which contains copper acetate and acetonitrile.[14, 46, 57] While the properties of the Cu-CNT wires improve over the reference CNT yarns, the absolute values of electrical conductivity of these composite wires are an order of magnitude lower than the SWCNT-Cu films shown in **Figure 8**; some of these studies also report interesting mechanical and current-carrying properties. The major difference between the small-scale film[14] and CNT-Cu wires is the type and length of CNTs used. It is not clear whether it is the SWCNTs that resulted in the superior electrical properties of the AIST paper[14] or the fact that CNT yarns are comprised of nanotube bundles that share several junctions with other tubes and bundles. The SWCNTs in the AIST study were highly porous, a few mm long, and not in the form of bundles (common for fibers and yarns). The reported conductivities in these samples were likely measured over millimeters length scale.

Most metals do not wet carbon, leading to poor interfacial bonding in metal-carbon composites. Two general approaches, both with negative consequences, have been tried to alleviate this problem. Nanocarbon surface can be functionalized to promote interactions with metals, however, functionalization degrades the properties of nanocarbons. While not suitable for electrical conductivity, mechanical properties of metal-carbon composites can benefit from partially compromised, functionalized nanocarbon properties for better bonding to the metals.[58] As an alternative, an intermediate metal can be added at the interfaces to improve the metal-carbon interactions. For example, while copper does not wet CNTs, there are a few metals such as nickel that both wet carbon and dissolve in copper.[59] Using such metals as an interfacial medium improves copper-carbon bonding.[59] Charge transfer between nanocarbon and metal is, however, altered due to the presence of the intermediate metal. The downside of this approach is that the intermediate metal dissolve in copper and significantly lower its conductivity.

Electrodeposition is specifically suitable for making Cu-graphene conductors. Negatively charged metallic precursors can coexist with GO to form homogeneous colloidal solutions. One step co-electrodeposition of graphene metal composite films becomes possible because both GO and metal reduction reactions can occur under cathodic conditions. Post-processing such as heat treatment (up to 3000ºC) in reducing atmospheres yields rGO with improved qualities, close to CVD grown graphene.[60] The addition of graphene to metals, specifically copper, can be performed via CVD. The quality of CVD graphene is much higher compared with rGO; however, the price is orders of magnitude higher and only limited loading of graphene in copper would be possible.

*2.2.3   Copper-Graphene Nanocomposites*

Extremely low graphene loadings (<100ppm) in copper has the potential to improve copper's conductivity. It seems that high temperatures and pressures are, however, needed for achieving an improved performance. It has been shown that the addition of ~20ppm graphene to copper, using a novel hot-extrusion process, improves the Cu-graphene conductivity to 103.5% IACS.[43] Unpublished work on a new process for Shear Assisted Processing and Extrusion (ShAPE[TM]) has also shown 105% IACS for copper enhanced with 10ppm of graphene. Most importantly, hot pressing several Cu foils with CVD grown graphene on them (less than 50ppm) has resulted in multi-layered Cu-graphene films exhibiting an *ultra-conductivity* of 117% IACS.[15] Two different mechanisms are proposed for the improved conductivity of Cu-graphene nanocomposites. i) The first one is healing of copper's defects under high pressures and temperatures due to the graphene presence. Most noticeably copper grains in the multi-layered Cu-graphene films grow in size and template onto the graphene films in a certain crystallographic direction, i.e., the (111). Electrical conductivity of single crystal copper has been reported to be as high as 112-114% IACS.[61] The application of elevated temperatures and pressures can further improve single crystal copper's conductivity to 117% IACS.[62] ii) The second mechanism is based on the contribution of graphene to conductivity. The conductivity of the embedded graphene in this scenario should be at least 3 orders of magnitude higher than copper, given its low loading (tens of ppm), to result in any meaningful increase in copper's conductivity. Electronic band structure of the graphene/copper interphase, in this case, is different than their individual band structures, yet showing the character bands of both graphene and copper. It is likely that these two mechanisms act together.[15]

In the field of advanced conductors, many interesting and unusual property enhancements have been realized for small scale samples but not always translated to the macroscale. The reader should, therefore, be cautiously optimistic about the promise of advanced conductors.

## 2.3   Characterization of advanced conductors

Engineers, materials scientists, chemists, and physicists conduct research on advanced electrical conductors, where each group performs numerous characterizations that are specific to their own field of study. Structural, chemical, mechanical, and physical properties that are of interest to one field may not be emphasized by another. Even if a research group is eager about a particular characterization, they may not have access to the equipment or the expertise to carry it out properly. It is obvious that almost any research effort in this field can benefit from inter- and multi-disciplinary teams to investigate different aspects of advanced conductors. Basic properties and structural features of interest in both nanocarbon-based and metal-carbon conductors are outlined and discussed in this section.

*2.3.1 Electrical Properties*

First and foremost, electrical conductivity and specific (per density) electrical conductivity of advanced conductors must be carefully characterized and reported; specific properties are desirable for weight critical applications. In terms of operation temperatures, advanced conductors are usually expected for use in the 300-500K (room to 200 ºC) temperature range and are therefore characterized at elevated temperatures. Cryogenic conductivity measurements (specially below 100K) can provide useful insights into the physics of conduction and structural defects present in the conductor.[63]

Characterization of conductivity and specific conductivity for a micro-scale fiber or thin film sample is rather difficult and complex, yet often overlooked. Extra care should be taken when conducting such measurements due to the usually small size of lab scale samples; see the section on uncertainty propagation below. To clarify this issue, let's consider electrical conductivity measurement of a copper wire and a CNT yarn. Copper wires have a tight diameter tolerance and are available in meters long form. One can easily measure the diameter of a copper wire and its conductivity over a long length, e.g., tens of centimeters. The uniformity of the diameter and lengthy samples result in small errors in measured electrical properties. Moreover, copper's density is well established and can be accurately measured; since relatively large amounts of copper wires exist for characterization. Specific properties of a copper wire can, therefore, be calculated with a high level of confidence; <1% difference with established values. In sharp contrast, a CNT yarn has a non-uniform (sometimes non-circular cross-section) diameter that is difficult to measure. The cross-section of each sample should be individually measured at several locations to achieve a statistically significant data. This is too challenging and researchers often measure as many points as possible under an electron microscope or using a non-contact micrometer. This approach can result in significant errors. For example, a 2 microns error (deviation) in dimeter measurements of a 40 microns diameter yarn, can result in a

~10% error in the reported electrical conductivity value. Measuring density of a CNT yarn is also highly challenging. First, a micro-balance or an ultra-microbalance is required to precisely measure the weight of the CNT yarn (usually samples are centimeters long and weighing tens or hundreds of µg). If a high-resolution balance is used, the earlier mentioned error in the diameter value can result in a ~10% error in density measurement; density is calculated in this case by dividing the yarn's mass by its volume calculated from yarn's length and cross-sectional area. If electrical conductivity is divided by the density, to achieve specific conductivity, the measurement error increases to ~20%. There is a way around this issue. Specific conductivity can be measured without knowing the cross-sectional area of a nanocarbon yarn. Conductivity and density are defined as $\sigma = L/RA$ and $\rho = M/LA$, respectively, where L and A are the wire length and cross-sectional area, and M is the wire mass. Specific conductivity is $\sigma/\rho = (L/RA)/(M/LA)$, which can be simplified to $\sigma/\rho = (L^2/RM)$.[27a]

*2.3.2 Ampacity*

Ampacity is an important property of a conductor that depends on the temperature-dependent electrical/thermal properties and geometry of the conductor.[64] For an insulated wire of a certain diameter, a higher electrical conductivity leads to a greater ampacity rating (higher current carrying capacity to reach a fixed temperature). The ampacity of advanced conductors is usually measured for a bare wire without any insulation. Detailed analysis for the measurement and modeling of ampacity for wires can be found elsewhere.[64]

*2.3.3 Mechanical Properties*

Basic mechanical properties of advanced conductors are often used to choose the right type and size conductor for an application. It is best to report all tensile properties of an advanced conductor, if tensile specimens can be fabricated. Tensile strength, modulus, failure strain (related to the maximum bending radius of a wire), and yield strength (for metal-nanocarbon conductors) are all used in the design and utilization of advanced conductors in different applications. Similar to the electrical conductivity, smallest deviations in dimensional measurements can translate to large errors in measured mechanical properties. Instead of measuring the strength and dividing it by the density, specific strength of a nanocarbon yarn can be measured as FL/M, where F is the failure load, L is the sample length, and M is the sample's mass; unit of specific strength is GPa/(gram/cm$^3$) or N/tex. Bending radius is another important mechanical property of a conductor, where it measures the minimum radius that a wire can be bent to repeatedly without braking.

*2.3.4 Nanocarbon Characterization*

Nanocarbons used in any study relating to advanced conductors must be fully and carefully characterized. This is due to the high sensitivity of properties and micro-structure in advanced conductors to the nanocarbon quality, composition, size, etc. The lack of such information makes it difficult to draw universal conclusions from these studies and even to reproduce their results. This is a very problematic issue, given that graphene or CNT from different suppliers or even batches from one supplier or made in a lab can have varied quality, functional groups, and impurity levels and compositions. It is, therefore, critical to report the type of CNT or graphene, their dimensions, defect density, dispersion quality (in a nanocomposite), and impurity levels. Comprehensive characterization of these properties is burdensome and unfortunately not provide in most studies.

Basic characterizations of CNT and graphene should include their Raman spectroscopy (assessing the quality and nature of nanocarbon), combined scanning electron microscopy (SEM) and energy-dispersive X-ray (EDX) spectroscopy (purity and morphology evaluation), and transmission electron microscopy (TEM) (assessing the quality and nanoscale features).[3, 21b] Several other spectroscopy and microscopy techniques can be used to further evaluate the composition, chemical groups, defects, size and morphology, and even electronic nature of nanocarbons.[3, 65]

*2.3.5 Metal-Carbon Nanocomposite Characterization*

The micro-structure of metal-carbon conductors should be characterized in the same manner pure metals have been traditionally studied. This includes metallographic studies on the grain size and orientation distributions (X-ray diffraction, etching and microscopy, and SEM/electron backscatter diffraction (EBSD)).[66] Examining the quality, state, and amount of nanocarbons in a metal nanocomposite is usually difficult.[51] Finally, since metal conductivity is highly sensitive to trace impurities, elemental analysis such as inductively coupled plasma (ICP)-based methods should be carried out. In addition to these essential characterizations, several other methods are available to further examine the structure of metals at different length scales.[66]

*2.3.6 Further notes on characterization of advanced conductors*

Comprehensive multi-scale characterization of structure, electrical, and mechanical properties of advanced conductors provides a better understanding of their potential for practical applications. For metal-carbon conductors, the basic characterizations include establishing the interrelationships between processing, microstructure, and electrical and mechanical properties. Some studies, however, lack such metallurgical studies of micro-structure (grain size distribution, texture, alloy concentrations) and measurements of mechanical properties and rather focus on nanoscale characterizations that are useful but secondary to the micro-structural ones.

Proper analysis of the experimental data is of utmost importance in advanced conductor research, especially for thin films and microscale fibers. In particular, uncertainty propagation should be taken into account in statistical analysis of experimental data.[67] A simple case demonstrating such analysis for reporting the measurement error for resistivity, $\rho$ (not to be confused with density), of a conductor wire of length L, resistance R, and cross sectional area A, is presented here. Resistivity is defined as **Equation 1**:

$$\rho = \frac{RA}{L} \qquad \text{Equation 1}$$

where R and L are measured independently, and A= $\pi D^2/4$. For uncertainty propagation analysis, the error, e, in resistivity measurement can be defined as **Equation 2**:

$$e_\rho = \sqrt{\left(\frac{\partial \rho}{\partial R}\right)^2 e_R^2 + \left(\frac{\partial \rho}{\partial A}\right)^2 e_A^2 + \left(\frac{\partial \rho}{\partial L}\right)^2 e_L^2} \qquad \text{Equation 2}$$

Similarly, error for the cross-sectional area is as **Equation 3**:

$$e_A = \left|\frac{dA}{dD}\right| e_D \qquad \text{Equation 3}$$

Replacing the following partial differentials in the $e_\rho$ equation results in **Equation 4**:

$$A = \frac{\pi}{4} D^2 \rightarrow e_A = \frac{\pi}{2} D e_D$$

$$\frac{\partial \rho}{\partial R} = \frac{A}{L} \ \& \ \frac{\partial \rho}{\partial A} = \frac{R}{L} \ \& \ \frac{\partial \rho}{\partial L} = -\frac{RA}{L^2}$$

$$e_\rho = \sqrt{\left(\frac{A}{L}\right)^2 e_R^2 + \left(\frac{R}{L}\right)^2 \left(\frac{\pi}{2} D e_D\right)^2 + \left(-\frac{RA}{L^2}\right)^2 e_L^2} \qquad \text{Equation 4}$$

The uncertainty in the measurements of the R, D, and L can be propagated to resistivity using the above formula.

## 2.4 Considerations for Utilizing Advanced Electrical Conductors

Approximate ranges for the specific and total electrical conductivity of advanced conductors, measured on a lab-scale, are plotted in **Figure 9**. Besides specific and total electrical conductivity, properties of interest for an electrical conductor are temperature coefficient of resistance (TCR), ampacity rating for a certain service temperature (usually imposed by the insulation material), maximum current carrying capacity in air or vacuum, thermal conductivity, mechanical strength and ductility, thermal expansion, creep resistance, corrosion resistance, and solderability. Advanced conductors outperform metals in many of these categories. Incorporating nanocarbons into metals can modify several of these properties as explained in this section.

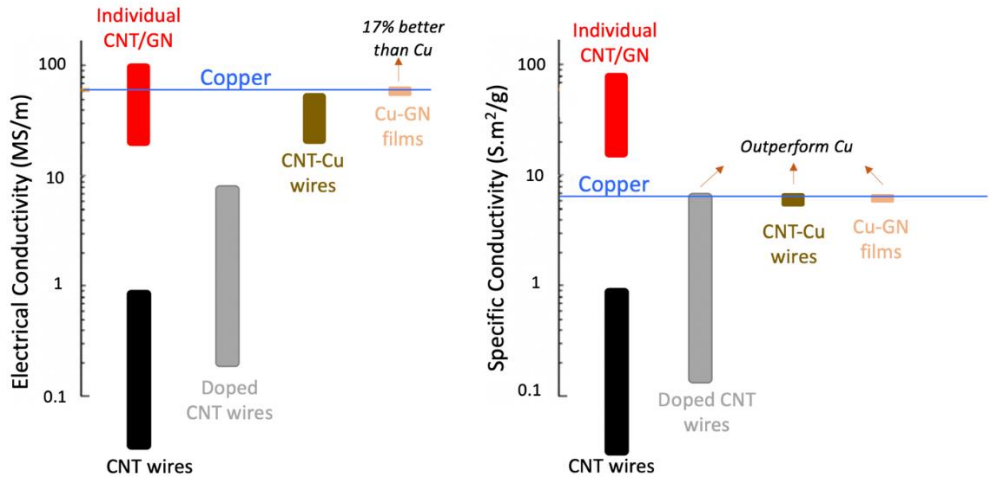

*Figure 9. Electrical properties of individual CNT, doped and pristine CNT wires, CNT-Cu, and Cu-graphene (GN) samples. Upper bounds for each plot represent measurement data from lab scale specimens.*

*Mechanical properties:* Besides a high strength, ductility and bendability are other mechanical properties that play important roles in the wire drawing process, wire installations, and design of electric machines. Strength of metals is readily enhanced by the addition of nanocarbons. Homogenous dispersion of nanocarbons in a metal matrix restrains dislocation movement and can potentially inhibit grain growth, both of which can improve the strength of the host metal.[68] Nanocarbon addition to Cu or Al, however, may adversely affect the ductility and bendability. Strength and bendability of CNT yarn conductors is several times better than Cu and Al. If the price can be justified, CNT yarns (tensile strength of 1-3 GPa) can be composited with copper or aluminum to achieve a conductor with superior mechanical performance and relatively high conductivity.[69]

*Thermal conductivity:* The high thermal conductivity of CNT and graphene (3000-6000 W/m/K) can be exploited to improve the conductivity of Cu, Al, and Ag (390, 200, and 400 W/m/K, respectively).[37, 56] Engineering nanocarbon-metal interfaces is required to minimize their interface thermal resistance.[70] An upper bound, **Figure 6**, for the thermal conductivity of metal-carbon composites can be calculated.

*TCR:* Superior conductivity at elevated temperature is of utmost importance for high-efficiency electric motors. Copper's high TCR results in a 50% increase in its resistivity at 150ºC, resulting in significant stator $I^2R$ losses[6d]. Improving copper's conductivity at temperatures above 100ºC can alone result in more efficient electrical motors and billions of dollars in electricity savings or increased range of battery-powered electric vehicles. There is one study exhibiting an elevated temperature conductivity in a CNT-Cu film that is better than copper. Such behavior has not been achieved yet for meters long nanocarbon-copper wires.

*Thermal expansion:* Graphene in its plane and CNTs along their longitudinal axis possess a negative CTE (-8/μ/K)[45] compared with the relatively large and positive CTE of copper and Al (23 and 17 /μ/K), and their addition to these metals can lower the overall CTE[45]. The reduction in CTE is commensurate with the volume fraction of the nanocarbon.

*Emissivity:* Nanocarbons have emissivity coefficients that are usually larger than that of Cu or Al. Covering these metals with a CNT or graphene coating can, therefore, increase their emissivity. For elevated temperature applications (above 100ºC) where the conductor surface is exposed (not like magnet wires), in addition to conduction and convection, the irradiation heat losses should be accounted for.[64] An increased emissivity in such cases can translate to a lower temperature (due to irradiation losses) for the nanocarbon-coated conductor, which is usually desired.

*Ampacity:* Ampacity is a term used to represent two somewhat different concepts. One is the ability of a conductor to carry a high current density without physically breaking, in either an ambient or low vacuum environment. Nanocarbons in copper have shown to inhibit electromigration, resulting in higher fusing temperatures/currents in high vacuum. The second one is the current that causes a conductor wire to reach a certain temperature under continuous operation; used to ensure the temperature of the wire remains below its service temperature. Ampacity rating for a certain temperature is difficult to improve, as it usually requires improving the conductor's electrical conductivity at the service temperature.[64]

*EMI Shielding:* EMI shielding relies traditionally on the use of high-density metals such as copper. Nanocarbon-based and metal-carbon conductors have shown to outperform conventional metal shielding and yet possess a lighter weight, resistance to corrosion, and flexibility. Both CNT and graphene have demonstrated excellent EMI shielding effectiveness.[37]

Discussions of any potential advanced conductor technology inevitably turn to scalable manufacturing, as one must be able to produce long lengths of any conductor for practical application. This is a necessary step towards a successful technology deployment, which depends on application-specific assessments of the needs for performance coupled with cost/benefit analysis that supports its use.

## 3 Research Challenges and Opportunities

### 3.1 The Numbers: Patents and papers

Analysis of the number of patents and papers and their country of origin tells us a great deal about the importance and state of any field. It also provides an idea of how much investment each country is dedicating to the field. Pure or doped nanocarbon (CNT or graphene) conductors are not expected to extensively replace copper or aluminum conductors. Nano-carbon metal composites, however, are more promising in terms of electrical conductivity and price, and one day can become ubiquitous. The analysis put forth in this section is, therefore, only for nanocarbon metal composites mentioning electrical conductivity as a characteristic. As shown in **Figure 10**, the number of journal papers on advanced metal-carbon conductors over the last fifteen years has seen an exponential growth. A plethora of work is published on mechanical properties of nanocarbon metal composites, however, their electrical conductivity has been less explored; possibly due to the degradation of metal's conductivity upon addition of nanocarbons via conventional processing methods. Analysis of the number of the patents since 2012, shows an overall growth trend on the number of published patents, on nano-carbon metal composites mentioning electrical conductivity as a characteristic; 30 in 2012, 35 in 2013, 42 in 2014, 71 in 2015, 83 in 2016, 94 in 2017, 134 in 2018, 118 in 2019. It is noteworthy that most of these patents are from China!

There will be more government initiatives on green energy in the coming years, especially with the increasing concerns with the global warming. The need to eliminate and reduce carbon foot print and machinery run on gas/diesel will force the transportation industry toward electro mobility in sea, air, and land. It may be argued that U.S. may be falling behind on a very important technological area, especially given that superior electrical conductors will play a significant role in the era of electro mobility. There will, of course, be many other benefits from replacing copper and aluminum with advanced conductors.

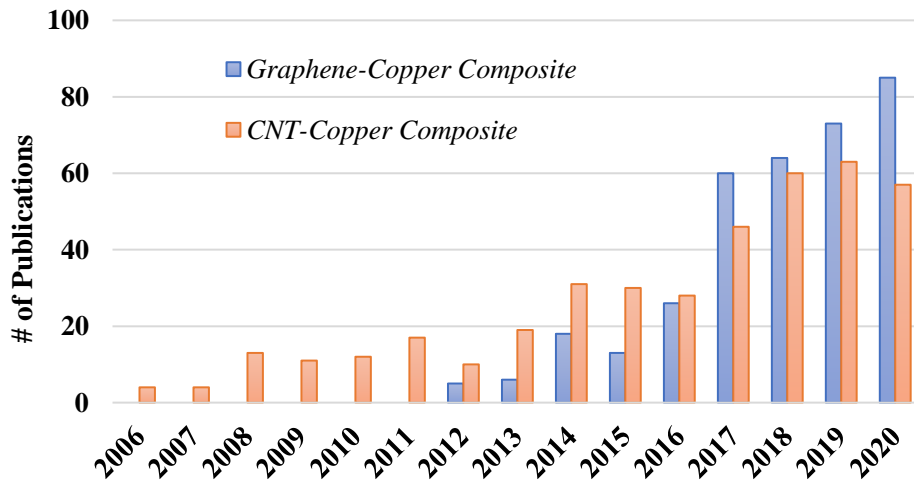

*Figure 10. Number of published articles by year found on Web of Science (Clarivate): keywords used for graphene-copper composite are "graphene" + "copper" + "electrical conductivity" or "resistivity" and for CNT-copper composite are "carbon nanotube" + "copper" + "electrical conductivity" or "resistivity"; 2020 numbers are estimated.*

## 3.2 A Brief History of Major Progress on Advanced Conductors

The U.S. government has supported basic and applied research on metal-carbon nanocomposites and nanocarbon-based advanced conductors over the last two decades. NASA and Airforce have been interested in lightweight conductors for data cables (satellite applications), EMI shielding, as well as elevated temperature wires for aircraft wiring and higher efficiency motors and transformers. Navy has mostly been interested in new power management solutions for energy dense systems in both aircraft and electric ships. Department of Energy's interest has been broader and mostly focused on electric motors where improvements to copper winding performance, especially at high temperatures, directly translates to more efficient motors and a reduced electricity bill, potentially saving billions of dollars. The less explored potential applications of nanocarbon-metal conductors are: i) pulsed magnets requiring wires with both a high conductivity and mechanical strength at cryogenic temperatures, ii) corrosion- and oxidation- resistant conductors, and iii) micro-electronics.

Over the last 15 years, there have been several Phase I SBIR/STTR awards (feasibility and commercial potential) on covetics but no phase II award so far. Most phase I efforts have failed to demonstrate meaningful bulk improvements in electrical, thermal, or mechanical properties of copper or aluminum covetics. Several fundamental challenges are yet to be resolved before the potential of covetics can be realized. There have been tens of phase I SBIT/STTR projects on carbon-nanotube based advanced conductors with a handful of those transitioning to phase II. The most successful stories for CNT-based materials are the EMI shielded co-axial data cables, as shown in **Figure 11**.[71] The EMI shields in satellites account for up to half of the copper harness weight and any reduction in their weight directly translates to significant savings in launching costs. There are two phase II Cu-CNT hybrid wire development projects, with the promise of a better than copper specific conductivity, current carrying capacity, and strength. Results from these projects are not published in the public domain at this time and their success is therefore not clear. A total of four projects on carbon nanotube-based advanced conductors for 150°C electric motors, with a similar funding and duration level to phase II SBR/STTR projects, have been funded by the ONR and DoE-EERE. To the best of author's knowledge, none of the conductors produced under these projects have surpassed the bulk electrical conductivity of copper at 150°C. Similar projects on ultra-conductive copper and CNT-based conductors have been funded in the Europe over the last 10 years. The major one was a multi-million Euro effort bringing in multiple universities and industries under a consortium to research the basic and applied science of advanced conductors. If one measures the outcome of these projects in terms of the fundamental science produced, then we have got a lot of 'bang for the buck'. The translation to commercialization, however, has been slow and breakthroughs in terms of physical property improvements have yet to occur. One can argue that superconductors have been also researched intensively over the last few decades and they also have failed to deliver their promise, only finding niche applications so far. As mentioned earlier, it took decades to scale-up the production of copper wire with uniform properties and patience may be the key to bringing the advanced conductors to fruition.[2]

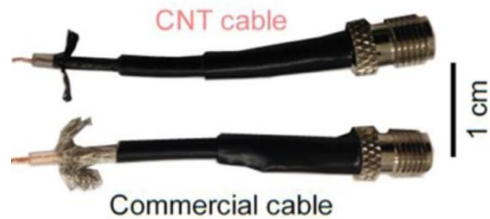

*Figure 11. A success story of CNT conductors as shielding material in co-axial data cables.[72]*

### 3.3 Future Research Directions

CNT-based conductors have been rigorously researched for over 20 years now,[56] but their promise of surpassing copper's electrical conductivity has not yet been realized.[14] The absolute room-temperature conductivity of Cu-CNT conductors is also lower than copper's. Although it has been seven years since the report of the lab scale Cu-CNT conductor with superb elevated temperature conductivity,[14] efforts to achieve similar properties on a wire scale have so far failed. Occasionally, doped-CNT and Cu-CNT conductors that outperform copper on a weight basis at room temperature are reported; aluminum's specific conductivity is twice that of copper's. The state-of-the-art in this field is advanced manufacturing of Cu-CNT conductors and synthesizing metallic SWCNTs and their assembly and doping into a functional conductor. A few groups, around the world, are working on the metallic-CNT-based conductors and if successful the scientific community will know about it soon.

Graphene has a relatively similar structure to CNTs, yet its interaction with Cu and Al is different. The addition of graphene to Cu has shown to affect copper's micro-structure and

conductivity.[60a, 73] Most notably, a very small loading of graphene (<0.001%) is required to cause measurable improvements in copper's conductivity. High temperature/pressure (950ºC and 50MPa) processing of several Cu-CVD graphene films has resulted in conductivities as high as 117% IACS at room temperature.[15]

Studies investigating the processing of Cu and graphene at temperatures below 750ºC and pressures less than10-20 MPa don't show improved physical properties of the resulting composites. It may be concluded that temperatures closer to the copper's melting point and higher pressures enhance copper's structure and Cu-graphene interfaces, rendering an overall improvement in the conductivity. The ShAPE™ process, developed by the Pacific Northwest National Lab (PNNL), has also shown to achieve up to 105% IACS for a ShAPE™ processed low quality graphene mixed with Cu.[43b] The process applies frictional shear, causing high temperatures, and pressures to form wires from a billet. Finally, there are controversial reports of improved conductivity in advanced conductors. Since these reports (e.g., claims of a ~200% IACS in a DWCNT-Cu or a conductivity of ~1000% IACS in a nitrogen-doped graphene on Cu) have not been independently verified, they are not cited here.

Maybe it is now graphene's turn and research should focus on graphene's potential for metal-carbon advanced conductors. We will know in years to come whether graphene can bring a revolution to the conductor industry. In addition to the current research in the field, the author has identified the following areas to be worthy of pursuing for the future:

- Cu-graphene and Al-graphene conductors have demonstrated encouraging properties and are worthy of further investigation. Specifically, processing of these nanocomposites under simultaneous high pressures and temperatures seems to induce interesting structural changes to the metal and metal-graphene interfaces. Future research directions in this area should focus on underpinning the science of conductivity improvement in metal-graphene conductors and utilizing this knowledge to realize scalable manufacturing of ultra-conductive metals.
- Improving mechanical strength or corrosion/oxidation resistance of Al alloys via graphene incorporation. Any improvements in this area can be impactful, specifically given that Al is relatively cheap and has a specific conductivity that is two times

- better than copper, yet some of its applications are limited by aluminum's low corrosion resistance and mechanical properties.
- The current R&D approach in the field of advanced conductors relies heavily on experiments. Theoretical/computational studies should be utilized to both understand and guide the experiments.
- A major challenge with any new materials innovation is its scalability from the lab scale to the industrial macroscale. Scaling lab scale technologies to reproducibly make ultra-conductors at the macroscale requires close collaboration between academia, government labs, and industry. Governments should initiate and support such efforts as venture investors tend not to fund early stage advanced conductor technologies.

## Disclaimer

## Acknowledgements

The author acknowledges the support by the International Copper Association (Copper Alliance®) and Office of Naval Research (ONR) under Grant No. N00014-18-1-2441. The author appreciates the constructive comments and discussions by Hal Stillman (International Copper Association), Malcolm Burwell (UltraConductive Copper Company), Keerti Kappagantula (Pacific Northwest National Laboratory), and Tina Kaarsberg (U.S. Department of Energy). The author thanks Pouria Khanbolouki for creating some of the schematics.